\documentclass[twocolumn,superscriptaddress,amsmath,amssymb,aps,prl]{revtex4-1}
\pdfoutput=1
\usepackage{amsmath}
\usepackage{amsfonts}
\usepackage{amssymb}
\usepackage{amsthm}
\usepackage{bm}
\usepackage{color}
\usepackage{dcolumn}
\usepackage{graphicx}
\usepackage{mathtools}
\usepackage{textgreek}
\usepackage[utf8]{inputenc}

\begin{document}
\title{Wetting behaviour of a three-phase system in contact with a surface}

\author{Biswaroop Mukherjee}
\email{b.mukherjee@sheffield.ac.uk}
\affiliation{Department of Physics and Astronomy, University of Sheffield, Sheffield S3 7RH, UK.}

\author{Buddhapriya Chakrabarti}
\email{b.chakrabarti@sheffield.ac.uk}
\affiliation{Department of Physics and Astronomy, University of Sheffield, Sheffield S3 7RH, UK.}

\date{\today}

\begin{abstract}
We extend the Cahn-Landau-de Gennes mean field theory of binary mixtures to understand the wetting thermodynamics of a three phase system, that is in contact with an external surface which prefers one of the phases. We model the system using a phenomenological free energy having three minima corresponding to low, intermediate and high density phases. By systematically varying the \textit{(i)} depth of the central minimum, \textit{(ii)} the surface interaction parameters, we explore the phase behavior, and wetting characteristics of the system across the triple point corresponding to three phase coexistence. We observe a non-monotonic dependence of the surface tension across the triple point that is associated with a complete to partial wetting transition. The methodology is then applied to study the wetting behaviour of a polymer-liquid crystal mixture in contact with a surface using a renormalised free energy. Our work provides a way to interrogate phase behavior and wetting transitions of biopolymers in cellular environments.
\end{abstract}
\maketitle

\section{Introduction} Wetting phenomena is ubiquitous in nature and arises in a variety of condensed matter systems ranging from classical fluids, to superconductors and Bose-Einstein condensates~\cite{p:indekeu.physica_A.v389.p4332.y2010,p:rolley.rmp.v81.p739.y2009,p:bonn.repprogphys.v64.p1085.y2001,p:degennes.rmp.v57.p827.y1985}. The most common example is a system having two bulk thermodynamic phases $\phi^{b}_{\alpha}(T)$, and $\phi^{b}_{\beta}(T)$, in contact with a surface that prefers one of them. For such systems, the wetting behavior can be understood using two equivalent formulations: \textit{i.e.} in terms of the \textit{(i)} contact angle $\theta$ describing the geometric profile of a sessile drop of two coexisting bulk phases at a temperature $T < T_{c}$, where $T_{c}$ corresponds to the bulk critical temparture, at the surface of a third phase~\cite{p:indekeu.physica_A.v389.p4332.y2010,p:bonn.repprogphys.v64.p1085.y2001,p:geogeghan.progpolysci.v28.p261.y2003}, and \textit{(ii)} profile $\phi(z)$, where $\phi$ corresponds to the concentration of $\alpha/\beta$ phase as a function of the distance $z$ from the surface of the third phase, which happens to be a spectator~\cite{p:cahn.jcp.v66.p3667.y1977}. In terms of the contact angle $\theta \rightarrow 0$, signals a transition from a partial to complete wetting, while in Cahn's approach, one has a macroscopic layer of one phase, $\phi_{\alpha}(T)$ in this case, residing at the surface, completely excluding the phase denoted by $\phi^{b}_{\beta}(T)$. A surface composition, $\phi_{s}$, intermediate between the densities of the two coexisting bulk phases, $\phi^{b}_{\alpha}(T)$ and $\phi^{b}_{\beta}(T)$ and decays smoothly to the bulk value of $\phi^{b}_{\beta}(T)$ is a characteristic of partial wetting. This transition from complete to partial wetting (also known as the interface unbinding transition) can be effected by lowering the temperature. Cahn showed that as one approaches the bulk critical point from below, the interfacial energy between the two phases goes to zero faster than the difference between their individual surface energies with the spectator phase. This thus necessitates a partial to complete wetting transition~\cite{p:cahn.jcp.v66.p3667.y1977} that has been well studied for small molecule mixtures.

The Cahn argument also applies to polymeric mixtures~\cite{p:pincus.jcp.v79.p997.y1983,p:i.schmidt.j_physique.v46.p1631.y1985,p:r.a.l.jones.prl.v66.p1326.y1991,p:r.a.l.jones.pre.v47.p1437.y1993,p:r.a.l.jones.polymer.v35.p2160.y1994,b:r.a.l.jones.1999,p:geogeghan.progpolysci.v28.p261.y2003}, however, there are two important differences. While for small molecule mixtures, the wetting transition occurs close to the bulk critical point, for polymer solutions, due to the low value of the interfacial tension between the immiscible phases, the transition occurs far from the bulk criticality~\cite{p:i.schmidt.j_physique.v46.p1631.y1985}. Secondly, unlike small molecule mixtures, one can study the wetting transition for polymers as a function of the molecular weights of the individual components. The complete to partial wetting transition is associated with lateral migration of material that results in interfaces being perpendicular to the confining wall~\cite{p:binder.Ann_rev_mat_res.v38.p123.y2008}. For a symmetric mixture of small molecules confined between asymmetric walls, (\textit{i.e.} where one wall preferentially attracts a phase) Parry and Evans~\cite{p:evans.prl.v64.p439.y1990} determined the concentration profile as a function of the confinement width and the temperature using mean field theory. This formulation was extended to polymeric fluids under symmetric~\cite{p:binder.Soft_Matt.v4.p1555.y2008} and asymmetric confinements~\cite{p:clarke.jcp.v131.p244903.y2009,p:clarke.jcp.v137.p174901.y2012}.

In all these situations, the bulk thermodynamics of the system is described by a mean field free energy with two minima corresponding to the stable phases of the system and a square gradient term which accounts for the free energy cost associated with spatial variations \cite{{b:r.a.l.jones.1999}}. The surfaces prefer one of the phases and is modelled by a surface free energy that depends on the local density at the wall. The problem of minimisation of the coupled bulk and the surface energies to obtain the concentration profiles can be mapped to a geometrical problem of Hamiltonian flow in phase space ~\cite{p:wortis.prb.v25.p3226.y1990}. Pandit and Wortis were the first to advocate the use of such phase portraits as a way of visualising the solutions of the wetting profiles obeying appropriate boundary conditions~\cite{p:wortis.prb.v25.p3226.y1990}.

While the above discussion describes wetting in binary mixtures of simple or polymeric fluids, whose bulk thermodynamics is dictated by a free energy with two stable minimum at temperatures below a bulk critical temperature, there are many important physical situations where  additional minima corresponding to locally stable phases may appear. Nematic ordering can induce additional local minimas in the 
free-energy landscapes as the anisotropic interactions are known to play an important role in the problem of 
polymer crystallisation \cite{p:r.larson_jcp_v150.p244903.y2019,p:l.fried_nat_mat_v5.p39.y2006}. 
Additionally, residual elastic interactions in the matrix arising from the presence of cross-links are known to severely modify free-energy landscapes of bulk mixtures and thus affect surface migration and wetting behaviour \cite{p:j.krawczyk.prl.v116.p208301.y2016,p:b.mukherjee.polymer.v12.p.1576.y2020}.
Other common examples are three phase systems \textit{e.g.} in polymer nematic mixtures \cite{p:matsuyama.epje.v9.p79.y2002,p.matsuyama.epje.v9.p89.y2002}, ternary amphiphiles~\cite{p:g.gompper_prl_v65.p1116.y1990,p:g.gompper_pra_v43.p3157.y1991,p:g.gompper_jcp_v102.p2871.y1994}, polymer-colloid mixtures~\cite{p:warren.epl.v20.p559.y1992,p:lekkerkerker.nuovcim.v64.p949.y1994} or metallic alloys  ~\cite{p:rost.natcomm.v8.p8485.y2015,p:manzoor.npjcompmater.v4.p47.y2018}. There has been a lot of recent interest in understanding  the wetting thermodynamics in ternary mixtures ~\cite{p:koga.faraday_disc.v146.p217.y2010,p:koga.prl.v104.p036101.y2010,p:koga.jphys_cond_matt.v28.p244016.y2016,p:koga.j_chem_phys.v150.p164701.y2019}. Depending on temperature and interaction parameters several possibilities exist, \textit{e.g.} one phase wets or spreads at the interface of the other two, or the three phases may meet along a line of common contact with three non-zero contact angles. The transition between these two states is an equilibrium, three-phase wetting transition and they appear in several varieties ranging from first to infinite order transitions~\cite{p:koga.prl.v104.p036101.y2010}.

In this paper we present a consistent mean-field treatment of the thermodynamics of wetting for a two-component, three-phase system, which is in contact with an external surface, that acts as a spectator. The free energy of such a class of systems is modelled by two order parameters (i) one distinguishing between the ordered and disordered phases and (ii) one that distinguishes between two disordered phases differing in density. We follow the Hamiltonian phase portrait method to understand wetting for such a model, using a renormalised free energy, obtained by integrating out the order parameter that distinguishes between the the high density disordered and ordered phases. The renormalised free energy is thus expressed in terms of a single order parameter corresponding to the relative density of the phases. We demonstrate that the stable solution for the surface fraction identified from the multiple solutions which appear in the Cahn construction, corresponds to the one which minimises the total surface free energy. We systematically vary the stability of the intermediate phase and the values of the surface interaction parameters and demonstrate the change in the nature of surface wetting transition as a result. Finally, we apply this scheme to study the wetting phase diagram of a model polymer dispersed liquid crystal \cite{p:matsuyama.epje.v9.p79.y2002,p.matsuyama.epje.v9.p89.y2002} described by a free energy which accounts for both phase separation between low and high density polymer phases and the nematic ordering of the liquid crystalline component. 

In the next section we present the basic framework of the wetting calculations, which is followed by a section on application of this method on the wetting transition in a simple binary polymer mixture. This is followed be a section on the wetting thermodynamics in the three-phase systems and in the final section we apply this formalism on a model polymer-nematic mixture.

\section{Wetting of a binary fluid in a semi-infinite geometry}
The basic aim of the wetting calculation is to minimise the total surface free-energy functional $\Delta G_{surf}(\phi_{s})$:
\begin{equation}
\Delta G_{surf}(\phi_{s}) = \Phi(\phi_{s}) + \int_{0}^{\infty} \left[ k(\phi) (\frac{d\phi}{dz})^2 + \Delta f'(\phi) \right] dz, \label{eqn_free_en_func}
\end{equation}
where $\Delta f'(\phi)$ is the bulk free energy contribution (after the common-tangent construction, see below), $k(\phi) (\frac{d\phi}{dz})^2$ accounts the free energy cost arising from spatial gradients of the order parameter $\phi$, with $k(\phi) = \frac{1}{36\phi(1-\phi)}$, and $\Phi(\phi_{s})$ \cite{b:r.a.l.jones.1999} accounts for the surface free-energy of the external surface located at $z=0$. The bulk free-energy, $\Delta f'(\phi)$, has a form that typically exhibits a single minimum at high temperatures, while it develops two distinct minima at lower temperatures, corresponding to two bulk thermodynamic phases. The thermodynamic equilibrium corresponding to the same chemical potential and osmotic pressure among the two coexisting thermodynamic phases is ensured by a common-tangent construction, 
\begin{eqnarray}
 \frac{\partial f}{\partial \phi} \bigg{\vert}_{\phi_{A}} = \frac{f(\phi_{B}) - f(\phi_{A})}{\phi_{B} - \phi_{A}} \nonumber \\
 \frac{\partial f}{\partial \phi} \bigg{\vert}_{\phi_{B}} = \frac{f(\phi_{B}) - f(\phi_{A})}{\phi_{B} - \phi_{A}}, \label{eqn_binodal}
\end{eqnarray}
where $\phi_{A}$ and $\phi_{B}$ are the two unknowns, which we identify as $\phi_{\alpha}(T)$ and $\phi_{\beta}(T)$, with the convention, $\phi_{\alpha}(T)$ $\leq$ $\phi_{\beta}(T)$. The free energy after the common tangent construction, 
\begin{equation}
\Delta f'(\phi) = f(\phi) - f(\phi_{\alpha}) - (\phi - \phi_{\alpha}) \frac{\partial f}{\partial \phi} \bigg{\vert}_{\phi_{\alpha}}, \label{grand_poten}
\end{equation} 
enters the subsequent wetting calculations (see Fig.~\ref{Fig_1}(a)).

The minimisation of the total free energy $\Delta G_{surf}(\phi_s)$ (see Eq.~\ref{eqn_free_en_func}) is done in two steps. First, the bulk contribution is minimised as a function of $\phi$ with the appropriate boundary conditions, \textit{i.e.} the local density at the external surface should be $\phi_s$. The functional form that minimises the bulk contribution expressed in terms of $\phi_s$ is then substituted back in Eq.~\ref{eqn_free_en_func}. As a result, $\Delta G_{surf}(\phi_s)$, the right hand side of Eq.~\ref{eqn_free_en_func} becomes a function of the yet undetermined surface fraction, $\phi_{s}$. This function is again minimised with respect to $\phi_s$ to obtain the surface fraction which then allows one to obtain the wetting profile. 

We use this framework to study wetting transition in a variety of systems. The equilibrium profiles, $\phi(z)$ which minimise the Lagrangian density, $L(\phi,\dot{\phi})$ (the integrand of Eq.~\ref{eqn_free_en_func}) obey the Euler-Lagrange equations
\begin{equation}
\frac{\partial \Delta f'}{\partial \phi} = 2k(\phi) \ddot \phi + \frac{\partial k}{\partial \phi}(\phi) \dot{\phi}^2, \label{eqn_Euler_Lagrange}
\end{equation}
where $\dot{\phi}$ = $\frac{d \phi}{dz}$ and $\ddot{\phi}$ = $\frac{d^{2} \phi}{dz^{2}}$. The Hamiltonian can be obtained from the Lagrangian via a Legendre transformation given by 
\begin{equation}
H(p,q) = p\dot{q} - L(q,\dot{q}) = \frac{p^2}{4k(q)} - \Delta f'(q), \label{eqn_hamiltonian}
\end{equation}
where the coordinate $q$ is $\phi$ and the conjugate momentum, $p$ given by 
\begin{equation}
p = \frac{\partial L}{\partial \dot{q}} = 2k(q)\dot{q}. \label{eqn_momentum}
\end{equation}

Since the Hamiltonian does not explicitly depend on $z$, it is a conserved quantity, which leads to the following equation
\begin{equation}
 k(\phi) \dot{\phi}^2 - \Delta f'(\phi) = A, \label{eqn_soln_gen}
\end{equation}
where the constant of integration $A = 0$, since in the bulk, both $\Delta f'(\phi)$ and $\dot{\phi}$ are zero. Thus the minimal solution is given by 

\begin{equation}
k(\phi) \dot{\phi}^2 = \Delta f'(\phi), \label{eqn_soln_specific}
\end{equation}
which implies the profile is given by,
\begin{equation}
\frac{d \phi}{dz} = \sqrt{\frac{\Delta f'(\phi)}{k(\phi)}}. \label{diff_eqn_profile}
\end{equation}

We take the positive sign  of the root of Eq.~\ref{diff_eqn_profile} if $\phi < \phi_{\infty}$, as is the case for all calculations outlined in this paper. Substituting this solution into Eq.~\ref{eqn_free_en_func}, allows us to change the integration variable from the spatial coordinate $z$ to the density $\phi$. As a result, we can rewrite Eq.~\ref {eqn_free_en_func} as, 
\begin{equation}
\Delta G_{surf}(\phi_{s}) = \Phi(\phi_{s}) + \int_{\phi_{s}}^{\phi_{\infty}} 2 \sqrt{\Delta f'(\phi) k(\phi)} d\phi. \label{eqn_surf_ene}
\end{equation}

In this work, we discuss a situation where the low density phase is preferred by the surface, \textit{i.e.}, $\phi_{s} < \phi_{\infty}$ and we take the positive sign of the above square root. For $\phi_{s}$ $>$ $\phi_{\infty}$ only $\Phi(\phi_{s})$ contributes to $\Delta G_{surf}(\phi_{s})$. In the final stage of the minimisation scheme, we minimise $\Delta G_{surf} (\phi_{s})$, given by Eq. \ref{eqn_surf_ene} with respect to $\phi_{s}$ to obtain the undetermined surface fraction. The surface free-energy used in this work is of the following form : $\Phi(\phi_{s}) = h \phi_{s} + \frac{1}{2}g \phi_{s}^2$, with $h < 0$ and $g > 0$. This choice makes the surface prefer a phase with $\phi_{s} = -h/g$. 

There are two ways to perform the final  minimisation, either by numerically computing $\Delta G_{surf} (\phi_s)$ for various values of $\phi_s$ and then finding its minima, or employing a Cahn-construction \cite{p:cahn.jcp.v66.p3667.y1977} by equating the first derivative of Eq.~\ref{eqn_surf_ene} with respect to $\phi_s$ to zero, yielding
\begin{equation}
 \frac{d \Phi (\phi_s)}{d \phi_s} = 2 \sqrt{\Delta f'(\phi_s) k(\phi_s)} = \Psi (\phi_s). \label{eqn_Cahn_constr}
\end{equation}
The surface fraction $\phi_s$, is then obtained from the intersection of the left and right hand side expressions of Eq.~\ref{eqn_Cahn_constr}, numerically, which can result in multiple solutions but the stable roots are found by comparison of areas. As discussed below both these procedures yield the same value of surface fraction $\phi_s$.

The profile is obtained by integrating Eq.~\ref{diff_eqn_profile}, which yields 
\begin{equation}
z = \int_{\phi_{s}}^{\phi(z)} \sqrt{\frac{k(\phi)}{\Delta f'(\phi)}} d \phi. \label{eqn_profile}
\end{equation}
The boundary condition is obtained by substituting $z = 0$ in Eq.~\ref{eqn_profile} and \ref{eqn_Cahn_constr} and taking their ratio, which finally yields,
\begin{equation}
2k(\phi_{s}) \frac{d \phi}{dz} \bigg{\vert}_{z = 0} = \frac{d \Phi(\phi_{s})}{d \phi_{s}}. 
\end{equation}

\section{Wetting behaviour of binary polymer mixtures}
As a simple example, we consider the complete to partial wetting transition, as the temperature $T$ is deceased, in a binary mixture consisting of long ($N_{A}$ = 100) and short ($N_{B}$ = 50) polymers, in presence of an external surface at $z = 0$, which prefers the short chain polymers (oligomers). The bulk thermodynamics is governed by a simple Flory-Huggins free energy \cite{b:rubinstein_colby_2003} of the form,
\begin{equation}
 f(\phi) = \frac{\phi}{N_{A}} \ln \phi + \frac{(1 - \phi)}{N_{B}} \ln (1 - \phi) + \chi \phi (1 - \phi), \label{polym_olig_free_en}
\end{equation}
where $\phi$ is the composition of polymers and $(1 - \phi)$ is the composition of the oligomers. The surface at $z = 0$ prefers the low $\phi$ component with the bare surface energy of the form 
\begin{equation}
\Phi(\phi_{s}) = h \phi_{s} + \frac{1}{2}g\phi_{s}^{2},  
\end{equation}
where $h < 0$ and $g > 0$ are the surface parameters. We choose $h = -0.00026$ and $g = 0.006$. This implies that $(1 - \phi)$, \textit{i.e.} the oligomer composition, is supposed to be high near this surface. The bulk phase of the polymer mixture becomes unstable when $\chi$ is increased beyond the spinodal value $\chi_{s}(\phi_{0})$, where $\phi_{0}$ is the composition of the initially uniform mixture. The value of the Flory-Huggins $\chi$ parameter at the spinodal is given by,
\begin{equation}
\chi_{s}(\phi_{0}) = \frac{1}{2} \left[ \frac{1}{N_{A} \phi_{0}} + \frac{1}{N_{B} (1 - \phi_{0})} \right], \label{spinodal_polym_olig}
\end{equation}

\begin{figure}[h]
\includegraphics[width=8.5cm]{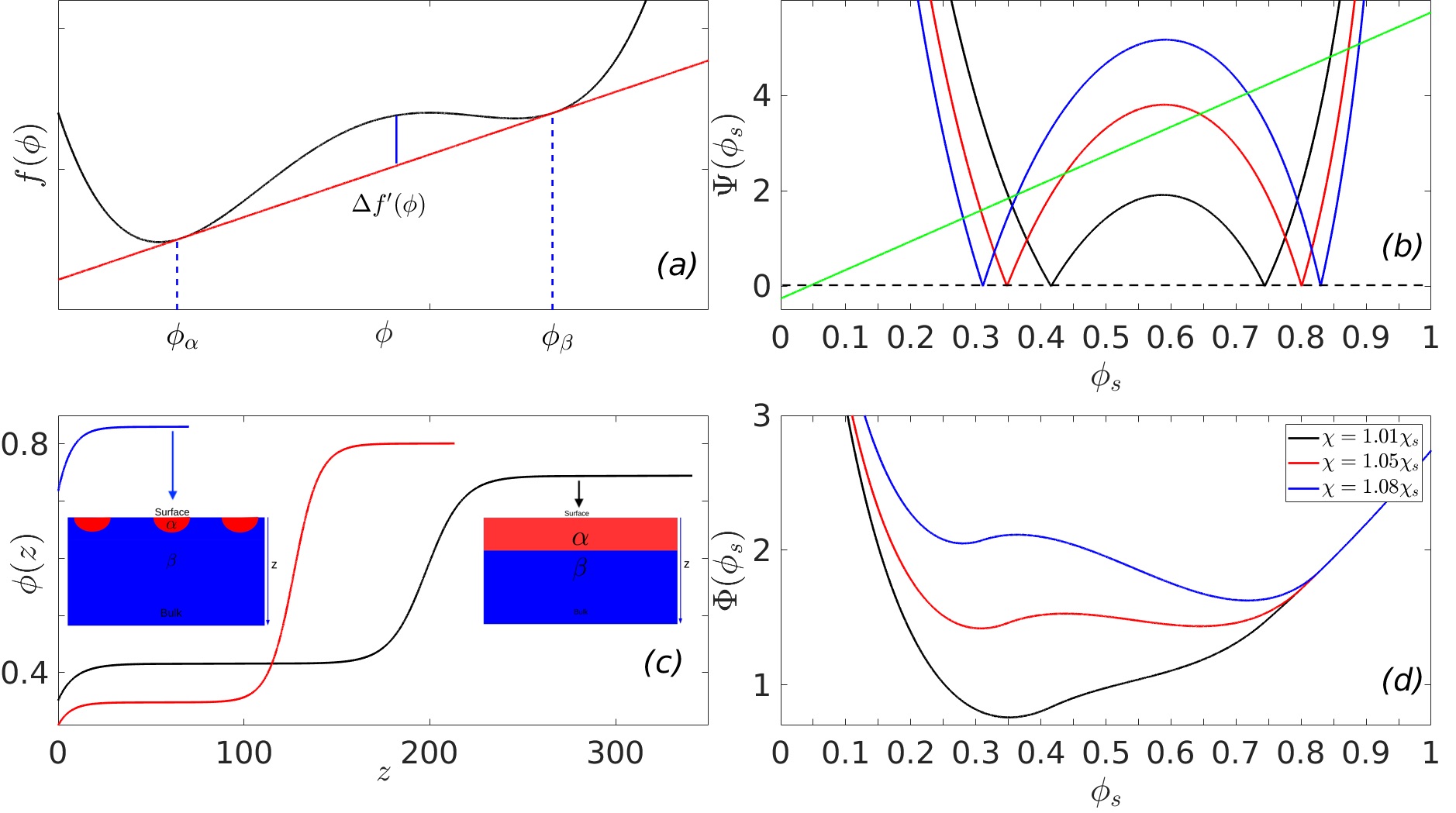} \\
\caption{The schematic double minimum free energy with common tangent in panel (a) (see Eq.~\ref{eqn_binodal}). The Cahn construction associated with the wetting calculation for the binary polymer mixtures with surface energy parameters $h=-0.00026$ and $g = 0.006$ is shown in panel (b). The concentration profiles for long polymers shows a complete to partial wetting transition in panel (c). The effective surface free energy, obtained after minimising the bulk thermodynamics of the system as a function of the surface fraction is shown in panel (d).} \label{Fig_1}
\end{figure}

Figure~\ref{Fig_1} shows the transition from complete to partial wetting in a binary polymer mixture, in contact with an external surface, as the immiscibility parameter $\chi$ is systematically increased (or the temperature of the system is decreased, since $\chi \propto 1/T$). Panel (b) shows the Cahn construction for the Flory-Huggins free energy for $\chi = 1.01, \text{(black)}, 1.05, \text{(red) and} 1.08 \chi_{s}(\phi_{0}) \text{(blue)}$ respectively, where $\chi_{s}(\phi_0)$ corresponds to the value of the immiscibility parameter at the spinodal (see Eq.~\ref{spinodal_polym_olig}). As shown the Cahn construction yields multiple solutions, and the surface fraction, $\phi_{s}$, is chosen for which $\Delta G_{surf}(\phi_{s})$ is minimum (see panel (d)). This procedure is consistent with the area rule used for choosing the stable solution \cite{p:bonn.repprogphys.v64.p1085.y2001}. The complete to partial wetting transition as the temperature is decreased is also evident from the change in the nature of the segregation profiles shown in Fig.~\ref{Fig_1} (c). At higher temperatures, \textit{i.e.} for $\chi = 1.01 \; \text{and} \; 1.05 \chi_{s}(\phi_{0})$, the low $\phi$ bulk phase (\textit{i.e.} oligomers) wets the external surface ($\phi_{s} < \phi_{\alpha}$) and completely expels the high density phase corresponding to polymers (see schematic in Fig.~\ref{Fig_1}(c)). When the temperature is decreased, \textit{i.e.}, for $\chi = 1.08 \chi_{s}(\phi_{0})$, a partially wetting profile, corresponding to  $\phi_{\alpha} < \phi_{s} < \phi_{\beta}$ is observed at the surface.

\section{Wetting in a three-phase system}
While the bulk thermodynamics of binary polymeric mixtures always involves a free-energy with two local minima occurring at bulk densities, $\phi_{\alpha}(T)$ and $\phi_{\beta}(T)$, complex mixtures with additional ordering fields, \textit{e.g.} ternary amphiphiles \cite{p:g.gompper_prl_v65.p1116.y1990,p:g.gompper_jcp_v102.p2871.y1994}, mixtures of nematics and polymers \cite{p:matsuyama.epje.v9.p79.y2002,p.matsuyama.epje.v9.p89.y2002} (we would be specifically discussing wetting in these systems later in this manuscript), can have free energies with additional metastable minima. The study of the influence of an ordering field on wetting transitions is very interesting with several technological applications in electro-optical devices~\cite{p:bunning.annurevmater.v30.p83.y2000,b:dahman.y2017} and high modulus fibres~\cite{b:zumer.y1996}. 

In this section, we extend the square-gradient mean field theory of wetting of a binary mixture to a three-phase system. In particular, we discuss the role of metastability on the wetting thermodynamics by studying a phenomenological form of free energy with three distinct local minima, whose location and relative depths can be varied. Since we do not have an explicit temperature dependent free energy, we study the wetting transitions as (i) a function of the stability of the central minimum and (ii) by varying the surface parameters, $h$ and $g$, which parametrizes $\Phi(\phi_{s})$, the interactions of the external wall with the system. We focus on the Cahn-construction for a ternary system and provide a criterion that dictates whether the wetting transitions are first order or continuous in nature. The three-phase free-energy that we consider has a piece-wise parabolic form,
\begin{equation}
f(\phi) = min \left[ f_{\alpha}(\phi), f_{\beta}(\phi), f_{\gamma}(\phi) \right], \label{ternary_free_en}
\end{equation}
where the $min$ function chooses the minimum of three individual functions given by,
\begin{eqnarray}
f_{\alpha}(\phi) = a_{\alpha}(\phi - \phi_{\alpha})^{2} + f_{\alpha}^{0} \nonumber \\
f_{\beta}(\phi) = a_{\beta}(\phi - \phi_{\beta})^{2} + f_{\beta}^{0} \nonumber \\
f_{\gamma}(\phi) = a_{\gamma}(\phi - \phi_{\gamma})^{2} + f_{\gamma}^{0}, \label{three_minima_free_en}
\end{eqnarray}
with the following set of parameters :  $\phi_{\alpha} = 0.1$, $\phi_{\beta} = 0.5$, and $\phi_{\gamma} = 0.9$, $a_{\alpha} = a_{\beta} = a_{\gamma} = 500$ and the relative heights of the three minima are set by $f_{\alpha}^{0} = 1$, $f_{\beta}^{0} = 3.5$, and $f_{\gamma}^{0} = 5$ respectively. 

\begin{figure}[h]
\includegraphics[width=8.5cm]{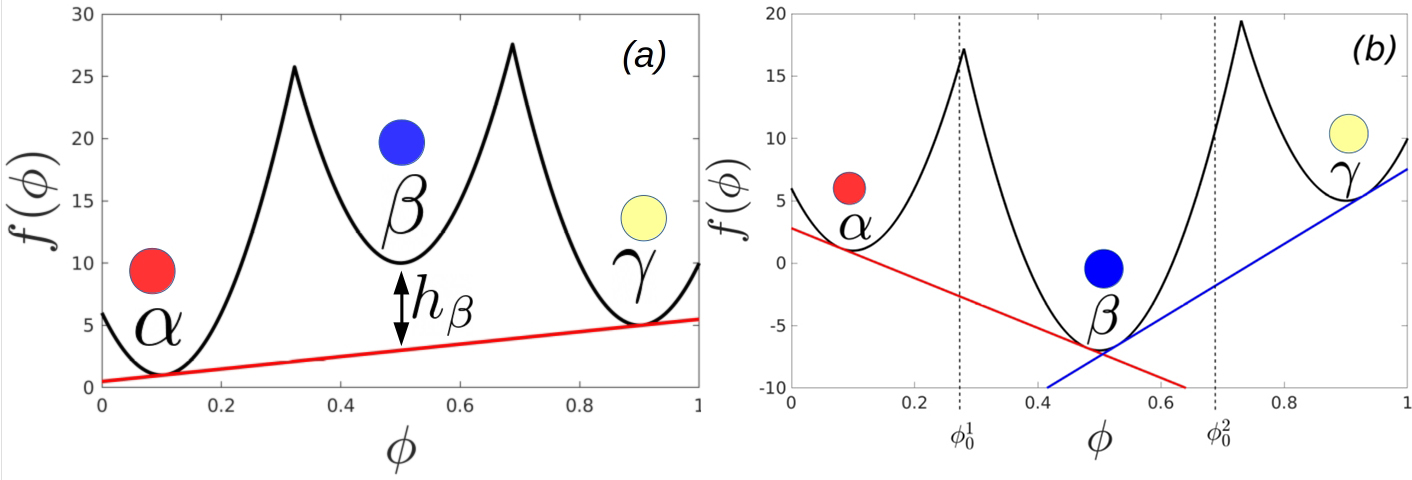} \\
\caption{The triple-minimum free energy used for the calculation. The low (red), intermediate (blue), and high density (yellow) phases correspond to densities  $\phi_{\alpha}$ = 0.1, $\phi_{\beta}$ = 0.5 and $\phi_{\gamma}$ = 0.9 respectively. The variable $h_{\beta}$ indicates the height of the barrier between the two thermodynamically stable phases between which the system splits.}
\label{triple_minim_free_en}
\end{figure}

\begin{figure}[h]
\begin{center}
\includegraphics[width=8.5cm]{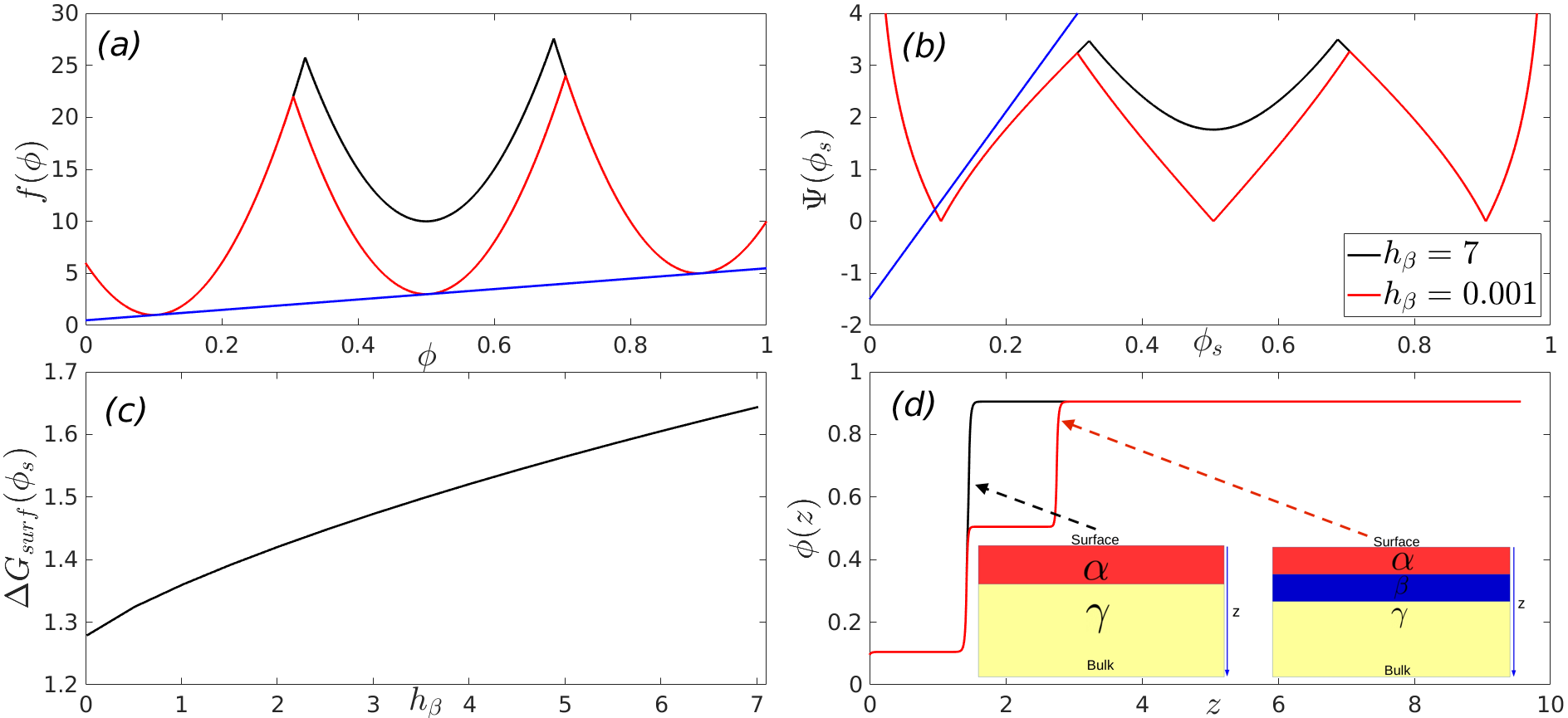} \\
\caption{The wetting thermodynamics as function of $h_{\beta}$, when it is positive and at the triple point. Panel (a) shows the free energies, panel (b) shows the Cahn constructions, panel (c) shows the dependence of the minimised surface free energy on $h_{\beta}$ and panel (d) shows the order parameter profiles.}
\label{wetting_h_beta_pos}
\end{center}
\end{figure}

We study the effects of the bulk thermodynamics on the wetting behaviour by systematically varying the free energy parameters corresponding to intermediate values of $\phi$ \textit{i.e.} $f^{0}_{\beta}$. As a result the depth of the central minimum $h_{\beta}$, (see Figure \ref{triple_minim_free_en}) is varied systematically by changing $f^{0}_{\beta}$, such that $-15 \leq f_{\beta}^{0} \leq 10$. The bare surface energy parameters are held fixed at $h = - 0.3 \mu_{bulk}$ and $g = -12 h$, where $\mu_{bulk}$ corresponds to the slope of the red line in Figure \ref{triple_minim_free_en} (a). Next, we study the wetting transition as a function of the surface parameters, \textit{i.e.} $h$ and $g$, close to the triple point (see red curve in Fig.~\ref{wetting_h_beta_pos}(a)). 

\begin{figure}[h]
\begin{center}
\includegraphics[width=8.5cm]{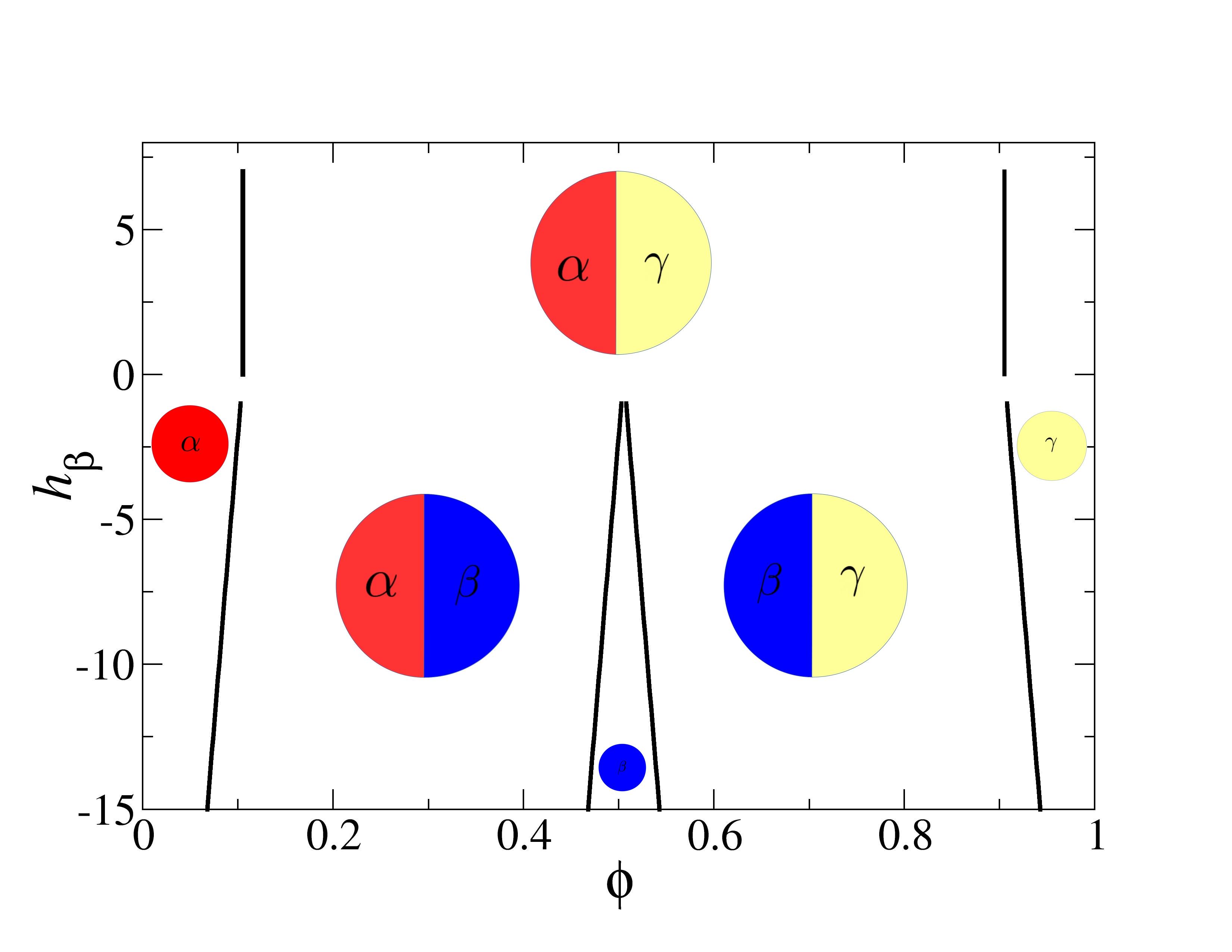} \\
\caption{The phase diagram for the three minimum free energy as a function of the stability of the phase $\beta$.}
\label{three_minim_phase_diag}
\end{center}
\end{figure}

The bulk phase diagram of the three-phase free energy as a function $h_{\beta}$ is shown in Figure \ref{three_minim_phase_diag}, where each region is designated by a colour of the phase/s that are stable in that region. For $h_{\beta} > 0$, the bulk free energy of a system, initially prepared with a uniform order parameter $\phi_{0}$, between $\phi_{\alpha}$ and $\phi_{\gamma}$, is minimised by splitting between these two minima in a manner which preserves the initial order parameter value of $\phi_{0}$. Thus the common tangent for the subsequent wetting calculation is drawn between the the minimum at $\phi_{\alpha}$ and $\phi_{\gamma}$ and the $\Delta f^{\prime} (\phi)$ for the subsequent wetting calculation should be constructed by subtracting 
off this common tangent from $f(\phi)$. Upon systematically decreasing $h_{\beta}$ a situation arises when the minima of all three parabolic free energies lie on a common tangent (Fig.~\ref{wetting_h_beta_pos}(a)). This is the triple point when the three phases coexist simultaneously.

For $h_{\beta} < 0$, \textit{i.e.} the $\beta$ minimum corresponds to the most stable phase. If the initial composition is such that $\phi_{0} < \phi_{\alpha}$, a single phase with composition $\phi_{\alpha}$ is chosen. When $\phi_{0}$  lies between the $\alpha$ and the $\beta$ minima, the bulk free energy is minimised by the system splitting between these two phases with the corresponding fractions following the lever rule \cite{b:rubinstein_colby_2003} and the $\Delta f^{\prime} (\phi)$ for the wetting calculation has been constructed by subtracting 
off this common tangent from $f(\phi)$. In this regime, the $\gamma$ component of the free energy does not enter the wetting calculations, as the $\alpha$ and the $\beta$ minimum have the lowest free energies according to our chosen parameters and hence the common tangent for the wetting calculation is drawn between these two states. The order parameter value, $\phi_{\infty}$, deep in the bulk is a value close to $\phi_{\beta}$. For higher values of the initial composition, $\phi_{0}$, $\beta$ phase becomes the most stable phase. Upon increasing $\phi_{0}$ further, the bulk free-energy would be minimised when the system splits between the $\beta$ and the $\gamma$ minimum and in this situation the order parameter value deep inside the bulk, $\phi_{\infty}$, would be close to $\phi_{\gamma}$. 
  
Figure \ref{wetting_h_beta_pos} shows the wetting thermodynamics as function of $h_{\beta} > 0$ and at the triple point, where the three phases coexist. Panel (a) shows the free energies corresponding to  $h_{\beta} = 7$ (black) and $h_{\beta} = 0.001$ (red) respectively. We assume that the initial composition, $\phi_0$  lies between $\phi_{\alpha}$ and $\phi_{\gamma}$. Thus the bulk free-energy is minimised by the system splitting appropriately between $\phi_{\alpha}$ and $\phi_{\gamma}$. We therefore draw a common tangent between these two minima, and the free energy, $\Delta f'(\phi)$, which enters the wetting calculation is obtained by subtracting this common tangent from the free energy $f(\phi)$ (see Eq.~\ref{grand_poten}). Panel (b) of Figure \ref{wetting_h_beta_pos} shows the corresponding Cahn constructions $h_{\beta} = 0.001, 7$ respectively. The derivative of the surface free energy, $d \Phi (\phi_{s})/d \phi_{s}$, (blue line in panel (b)) intersects the curve $2 \sqrt{\Delta f'(\phi_s) k(\phi_s)}$ (RHS of Eq.~\ref{eqn_Cahn_constr}), only at one point, which yields the surface fraction, $\phi_{s} < 0.1$. The equilibrium value of the high-density phase, corresponds to the material concentration deep in the bulk, $\phi_{\infty} \approx 0.9$. Thus, these parameters, set the lower and upper limits of integration for the expressions appearing in Eq.~\ref{eqn_surf_ene} and \ref{eqn_profile}. 

Panel (c) shows the monotonically decreasing minimised surface free energy (the minimum of $\Delta G_{surf}(\phi_{s})$), or the surface tension, as a function of $h_{\beta}$ and panel (d) shows the order parameter profiles. From Eq. \ref{eqn_surf_ene} it is clear that the surface tension has two contributions, one arising from the bare surface energy and the second from the area under the curve, $2 \sqrt{\Delta f'(\phi_s) k(\phi_s)}$. In this case the surface fraction, $\phi_{s}$ is independent of the variation in $h_{\beta}$, thus, while the bare surface energy remains unchanged the area under the curve, $2 \sqrt{\Delta f'(\phi_s) k(\phi_s)}$, monotonically decreases with $h_{\beta}$. This leads to the monotonic decrease in surface tension with $h_{\beta}$. Similar behaviour has also been observed in calculations of surface tension in bulk systems with multiple minimas in the free energy landscape \cite {p:bagchi.jpcb.v117.p13154.y2013}. It is clear from panel (d) that away from the triple point, when $h_{\beta}$ is positive and high, the order-parameter profile starts from $\phi_{s} <$ 0.1 ($\alpha$ phase) and finally tends to its value of $\phi_{\infty}$ $\sim$ 0.9 ($\gamma$ phase) and the effect of the meta-stable $\beta$ phase is negligible. Close to the triple point (see panel  (d) of Figure \ref{wetting_h_beta_pos}) there is a split interface with the surface wet by the $\alpha$ phase thereby completely excluding the $\beta$ and the $\gamma$ phases from the surface. The $\alpha$ phase is then wet by the $\beta$, which in turn is wet by the $\gamma$ phase as one moves from the surface to the bulk. Schematic order parameter configurations for these two situations are shown in the insets in panel (d) of Figure \ref{wetting_h_beta_pos}. 

\begin{figure}[h]
\begin{center}
\includegraphics[width=8.5cm]{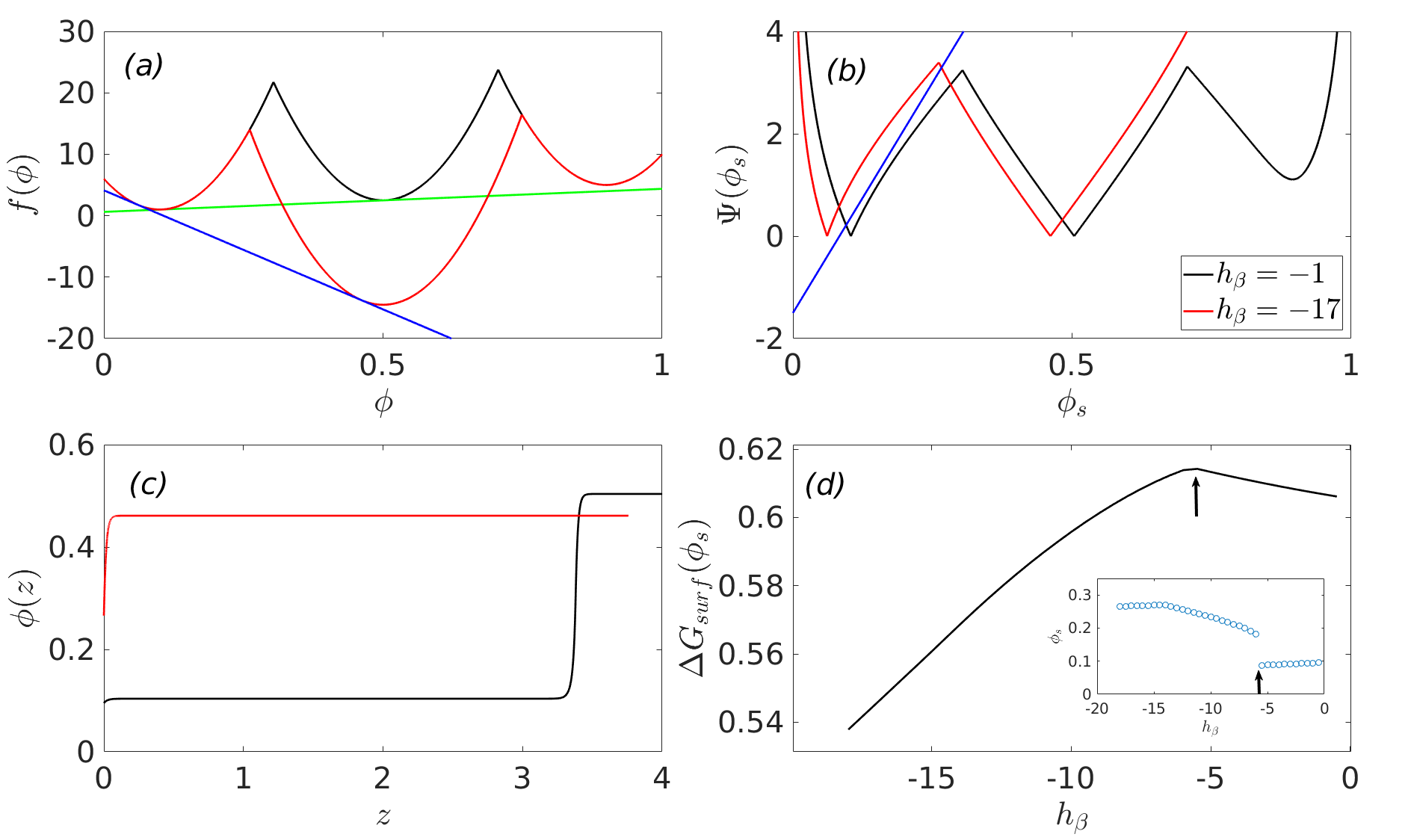} \\
\caption{The wetting thermodynamics as function of $h_{\beta}$, when it is negative and when the $\alpha$ and the $\beta$ phases coexist. Panel (a) shows the free energies, panel (b) shows the Cahn constructions, panel (c) shows the segregation profiles and 
panel (d) shows the dependence of the minimised surface free energy on $h_{\beta}$. The inset to panel (d) shows the dependence of the 
surface fraction $\phi_{s}$ on $h_{\beta}$, which signifies a transition from complete to partial wetting as one decreases $h_{\beta}$.}
\label{wetting_alpha_beta_h_beta_neg}
\end{center}
\end{figure}

Figure \ref{wetting_alpha_beta_h_beta_neg} summarises the thermodynamics of wetting as a function of $h_{\beta}$ when it is negative and the initial composition, $\phi_{0}$, of the system is bracketed by $\phi_{\alpha}$ and $\phi_{\beta}$ (see Figure \ref{three_minim_phase_diag} and the composition $\phi_{0}^{1}$ marked in Figure \ref{triple_minim_free_en} (b)). Panel (a) of Figure \ref{wetting_alpha_beta_h_beta_neg} shows the free-energies at two representative values of h$_{\beta}$ and the common tangents constructed between the free-energy minimum corresponding to $\phi_{\alpha}$ and $\phi_{\beta}$. Thus the relevant free energy $\Delta f'(\phi)$, which enters the wetting calculation is obtained by subtracting this common tangent from the free energy $f(\phi)$ shown in panel (a) of Figure \ref{wetting_alpha_beta_h_beta_neg}. As a result, the value of the order parameter deep inside the bulk would be $\sim$ $\phi_{\beta}$ = 0.5. As $h_{\beta}$ becomes increasingly negative the value of $\phi$, in the vicinity of $\phi_{\alpha}$, at which the common tangent between the $\alpha$ and the $\beta$ minimum intersects the free energy $f(\phi)$, decreases. This leads to an interesting behaviour in the wetting phenomena. Panel (b) shows the Cahn construction for determining the surface fraction. The location where the line corresponding to $\frac{d \Phi (\phi_s)}{d \phi_s}$ (blue line in panel (b)) becomes positive occurs at $\phi_{s} = -h/g$. For small absolute values of $h_{\beta}$, the value of $\phi_{s}$ at which $2 \sqrt{\Delta f'(\phi_s) k(\phi_s)}$ becomes zero (or $\Delta f'(\phi_s)$ becomes zero) is greater than $\phi_{s} = -h/g$. This signifies a complete wetting of the surface by the $\alpha$ phase as shown in the order parameter profile, black line in panel (c). As $h_{\beta}$ becomes increasingly negative, a situation arises when the value of $\phi_{s}$ at which $\Delta f'(\phi_s)$ becomes zero, is less than $\phi_{s} = -h/g$ and this leads to a transition from complete to partial wetting and the red line in panel (c) yields a profile where the surface is partially wet by both the $\alpha$ and the $\beta$ phases. This transition from complete to partial wetting results in a non-monotonic dependence of the surface tension or the minimised surface free energy, $\Delta G_{surf}(\phi_{s})$, shown in panel (d) of Figure \ref{wetting_alpha_beta_h_beta_neg}. The value of $h_{\beta}$ at which the non-monotonic behaviour in $\Delta G_{surf}(\phi_{s})$ arises, is that value where a transition from complete to partial wetting, of the surface by the $\alpha$ phase, occurs. This is shown in in the inset of panel (d), which shows the dependence of the surface fraction, $\phi_{s}$ on $h_{\beta}$. This dependence of the surface tension is unlike what had been observed in the situation when $h_{\beta}$ was positive.

\begin{figure}[h]
\begin{center}
\includegraphics[width=8.5cm]{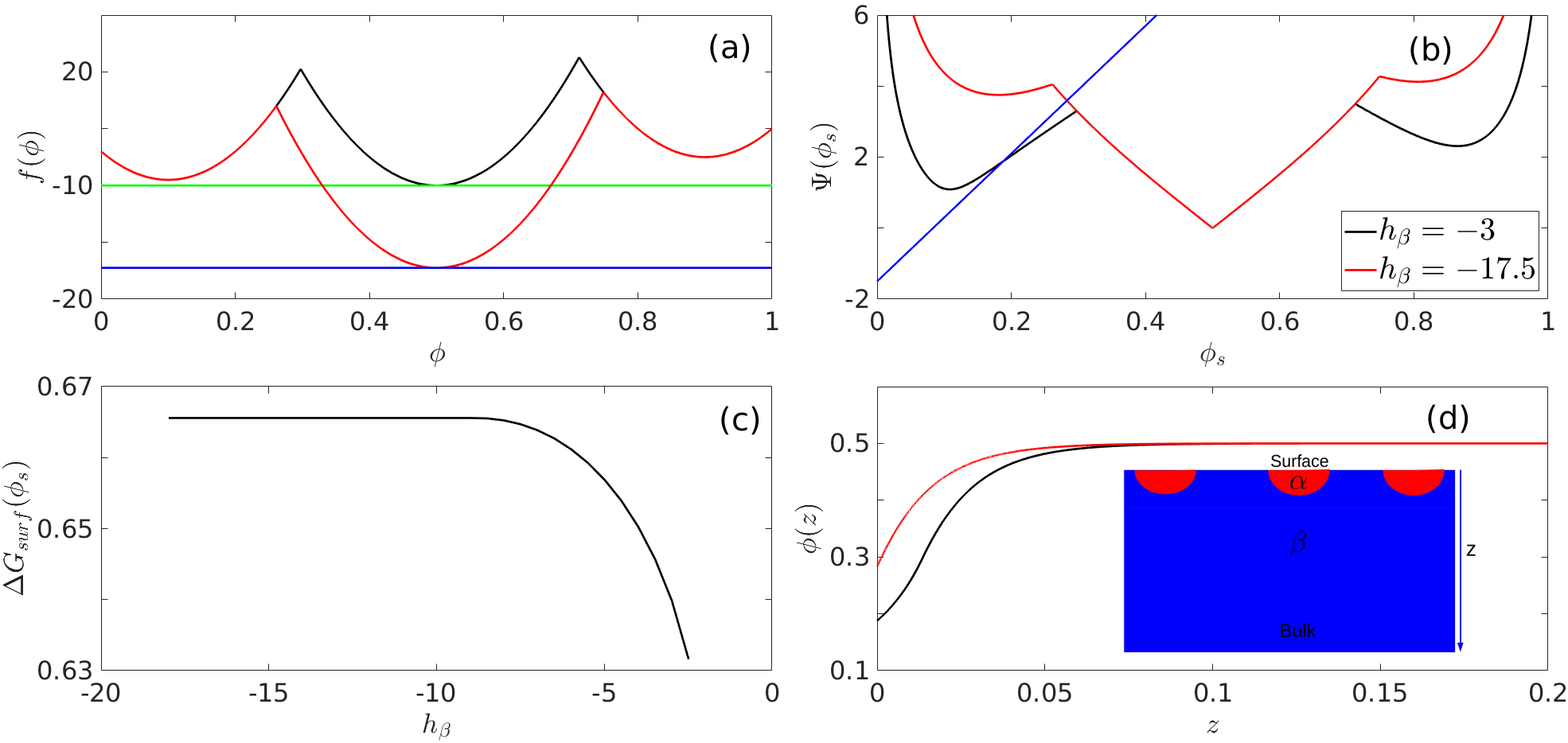} \\
\caption{The wetting thermodynamics as function of $h_{\beta}$, when it is negative and the $\beta$ phase is the most stable one. Panel (a) shows the free energies, panel (b) shows the Cahn constructions, panel (c) shows the dependence of the minimised surface free energy on $h_{\beta}$ and panel (d) shows the segregation profiles.}
\label{wetting_beta_h_beta_neg}
\end{center}
\end{figure}

For negative $h_{\beta}$, and $\phi_{0} \sim \phi_{\beta}$, the $\beta$ minimum is the only stable state available, which minimises the free energy of the system. In this situation the reconstructed free-energy for the wetting calculation is obtained by drawing a horizontal tangent to the full free-energy at $\phi_{\beta}$ and subtracting this line from $f(\phi)$. The summary of the wetting calculation in this regime is presented in Figure \ref{wetting_beta_h_beta_neg}, where panel (a) shows the free-energies and the horizontal tangent for two chosen values of $h_{\beta}$. Panel (b) of Figure \ref{wetting_beta_h_beta_neg} shows the Cahn plots for obtaining the surface fraction and in these situations there is only one intersection between the red and black bulk contributions of $2 \sqrt{\Delta f'(\phi_s) k(\phi_s)}$ and the surface contribution arising from the $\frac{d \Phi (\phi_s)}{d \phi_s}$ term is shown in blue. With decreasing $h_{\beta}$, the value of surface fraction $\phi_{s}$ systematically increases (see the Cahn plots in panel (b) of Figure \ref{wetting_beta_h_beta_neg}). Thus, in this situation, the two terms contributing to the surface tension in Eq. \ref{eqn_surf_ene} has opposite dependence with decreasing $h_{\beta}$. While the bare surface energy increases with $\phi_{s}$ the area under $2 \sqrt{\Delta f'(\phi_s) k(\phi_s)}$ decreases, with the bare surface energy contributing more and this leads to the initial increase in the surface tension with decreasing $h_{\beta}$ (see panel (c)). Once $h_{\beta}$ falls below $h_{\beta} \lesssim -8$, the surface line in panel (b) moves from the parabola corresponding to $\alpha$ minimum to the one corresponding to the $\beta$ minimum. After this point the surface fraction remains invariant upon further decrease of $h_{\beta}$ and as a result the surface tension in panel (c) also shows a plateau. Panel (d) of Figure \ref{wetting_beta_h_beta_neg} shows the segregation profiles for two values of $h_{\beta}$ and in both these situations one observes partial wetting and the inset shows a two-dimensional, schematic representation of the order parameter profile. \\

If the initial composition, $\phi_{0}$, is bracketed by $\phi_{\beta}$ and $\phi_{\gamma}$, there are two possibilities for minimising the bulk free energy, either (a) the $\phi_{0}$ is divided between the $\beta$ and the $\gamma$ minimum by order-parameter conservation and the minimum free energy is given by $F_{A}$ or (b) the system tries to minimize its free-energy by splitting into the three minimas conserving the order parameter and the minimum free energy is given by $F_{B}$. This second possibility arises as $f(\phi_{\alpha}) < f(\phi_{\gamma})$. We prove below that $F_{A}$ is always less than $F_{B}$, which means that an initial uniform composition, $\phi_{0}$, which is between $\phi_{\beta}$ and $\phi_{\gamma}$, will always be split into order-parameter values obtained by drawing a common tangent between the $\beta$ and the $\gamma$ minimum. The free-energy $F_{A}$ is given by 
\begin{equation}
F_{A} = \frac{\phi_{\gamma} - \phi_{0}}{\phi_{\gamma} - \phi_{\beta}} f(\phi_{\beta}) + \frac{\phi_{0} - \phi_{\beta}}{\phi_{\gamma} - \phi_{\beta}} f(\phi_{\gamma}).
\end{equation}
We explore the possibility of lowering the free energy of the system by splitting it into three minima. Assuming the mass conservation constraint,
\begin{equation}
    \phi_{0} = f_{\alpha} \phi_{\alpha} + f_{\beta} \phi_{\beta} + (1 - f_{\alpha} - f_{\beta}) \phi_{\gamma}.
\end{equation}
The above equation allows us to express the fractions $f_{\alpha}$  and $f_{\gamma}$ in terms of the fraction $f_{\beta}$,
\begin{eqnarray}
    f_{\alpha} = \frac{(\phi_{\gamma} - \phi_{0}) - f_{\beta}(\phi_{\gamma} - \phi_{\beta})}{(\phi_{\gamma} - \phi_{\alpha})}  \nonumber \\
    f_{\gamma} = \frac{(\phi_{0} - \phi_{\alpha}) - f_{\beta}(\phi_{\beta} - \phi_{\alpha})}{(\phi_{\gamma} - \phi_{\alpha})}.
\end{eqnarray}

From the above fractions one can write the free energy, where the initial order parameter has been partitioned into the three free energy minimum, in the following form,

\begin{eqnarray}
    F_{B} &=& \left[ \frac{(\phi_{\gamma} - \phi_{0}) - f_{\beta}(\phi_{\gamma} - \phi_{\beta})}{(\phi_{\gamma} - \phi_{\alpha})} \right]f(\phi_{\alpha}) + \nonumber \\ &f_{\beta} f(\phi_{\beta})& + \left[ \frac{(\phi_{0} - \phi_{\alpha}) - f_{\beta}(\phi_{\beta} - \phi_{\alpha})}{(\phi_{\gamma} - \phi_{\alpha})} \right]f(\phi_{\gamma})
\end{eqnarray}

It is evident from the above expressions that $f_{\gamma} > f_{\alpha}$, owing to the choice of parameters for our model free-energy, and both of them linearly decrease as one increases $f_{\beta}$, due to the constraint that their sum should be equal to unity. Thus upon systematically increasing $f_{\beta}$, $f_{\alpha}$ reaches zero first and this occurs when $f_{\beta} = \frac{\phi_{\gamma} - \phi_{0}}{\phi_{\gamma} - \phi_{\beta}}$ and 
$f_{\gamma} = \frac{\phi_{0} - \phi_{\beta}}{\phi_{\gamma} - \phi_{\beta}}$. At this point the free energy of the system is $F_{A}$ and thus this proves that $F_{B}$ cannot be lower than $F_{A}$, implying that when $\phi_{\beta} < \phi_{0} < \phi_{\alpha}$, the lowest free energy would be obtained by splitting between $\beta$ and the $\gamma$ minimum. This thus implies that the relevant common tangent must be between  
the free-energy minimum at $\phi_{\beta}$ and $\phi_{\gamma}$ and the $\Delta f^{\prime} (\phi)$ should be constructed by subtracting 
off this common tangent from $f(\phi)$.

\begin{figure}[h]
\begin{center}
\includegraphics[width=8.5cm]{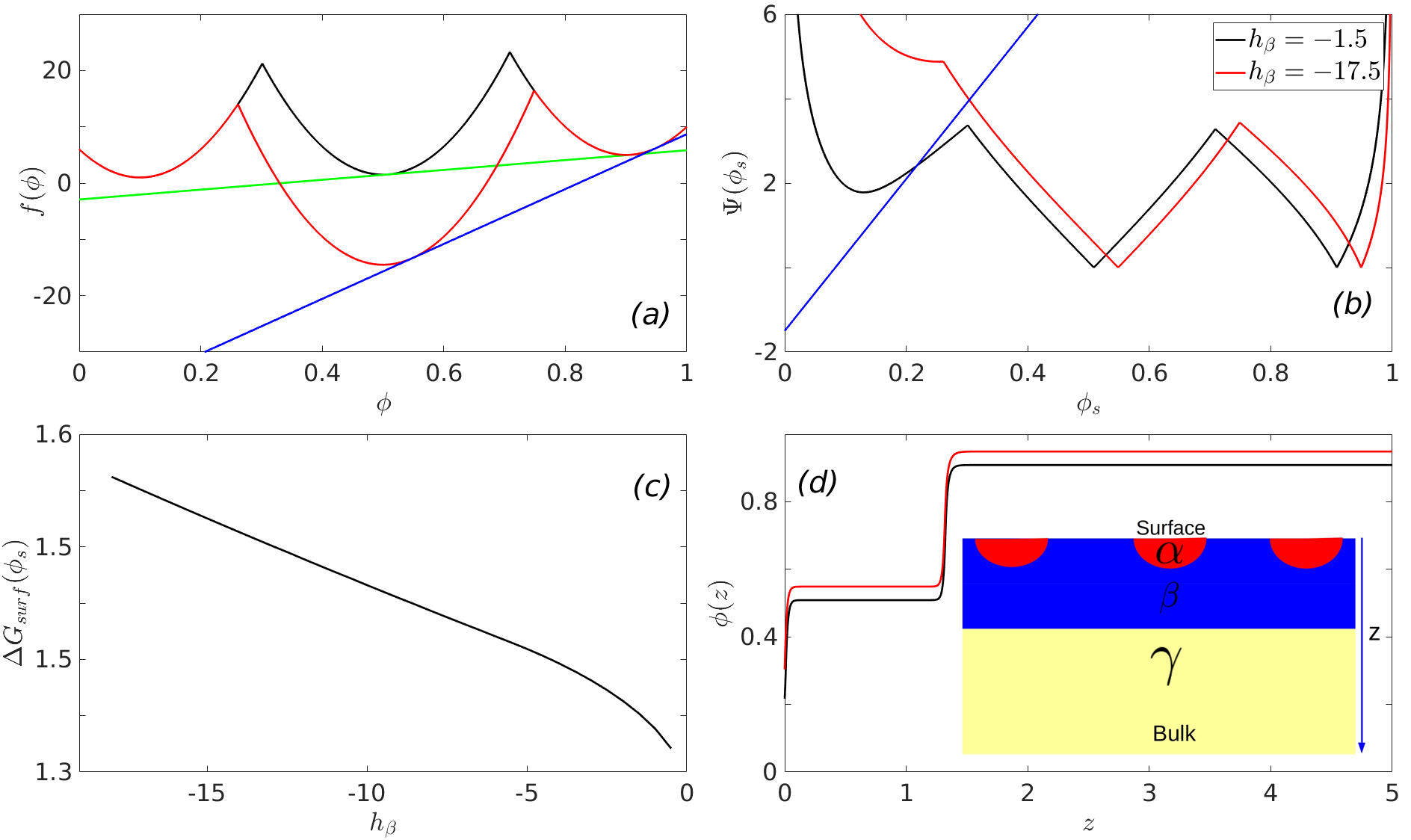} \\
\caption{The wetting thermodynamics as function of $h_{\beta}$, when it is negative and when the $\beta$ and the $\gamma$ phases coexist. Panel (a) shows the free energies, panel (b) shows the Cahn constructions, panel (c) shows the dependence of the minimised surface free energy on $h_{\beta}$ and panel (d) shows the segregation profiles.}
\label{wetting_beta_gamma_h_beta_neg}
\end{center}
\end{figure}

Figure \ref{wetting_beta_gamma_h_beta_neg} summarises the wetting thermodynamics for negative h$_{\beta}$, when the initial composition $\phi_{0}$, is split between the $\phi_{\beta}$ and $\phi_{\gamma}$ minimum (the composition $\phi_{0}^{2}$ in Figure \ref{triple_minim_free_en} (b)). Panel (a) of Figure \ref{wetting_beta_gamma_h_beta_neg} shows the free-energies and the common tangents, panel (b) shows the Cahn plots yielding the surface fraction, $\phi_{s}$. Panel (c) shows the variation of the minimised surface free energy as a function of the decreasing $h_{\beta}$ and panel (d) shows the segregation profiles for two values of $h_{\beta}$. The inset to panel (d) shows a schematic, two-dimensional order parameter profile, which signifies that the surface is 
partially wetted by both $\alpha$ and $\beta$ phases. In this situation, the minimised surface free energy, $\Delta G_{surf}(\phi_{s})$, increases with decreasing $h_{\beta}$. This can be physically understood from the fact that the bare surface free energy is minimum for $\phi_{s} \sim$ 0.083 and it increases for 
higher values of $\phi_{s}$. With decreasing $h_{\beta}$, the value of $\phi_{s}$ increases, thus it leads to a monotonic increase of the total surface free energy.

\begin{figure}[h]
\begin{center}
\includegraphics[width=8.5cm]{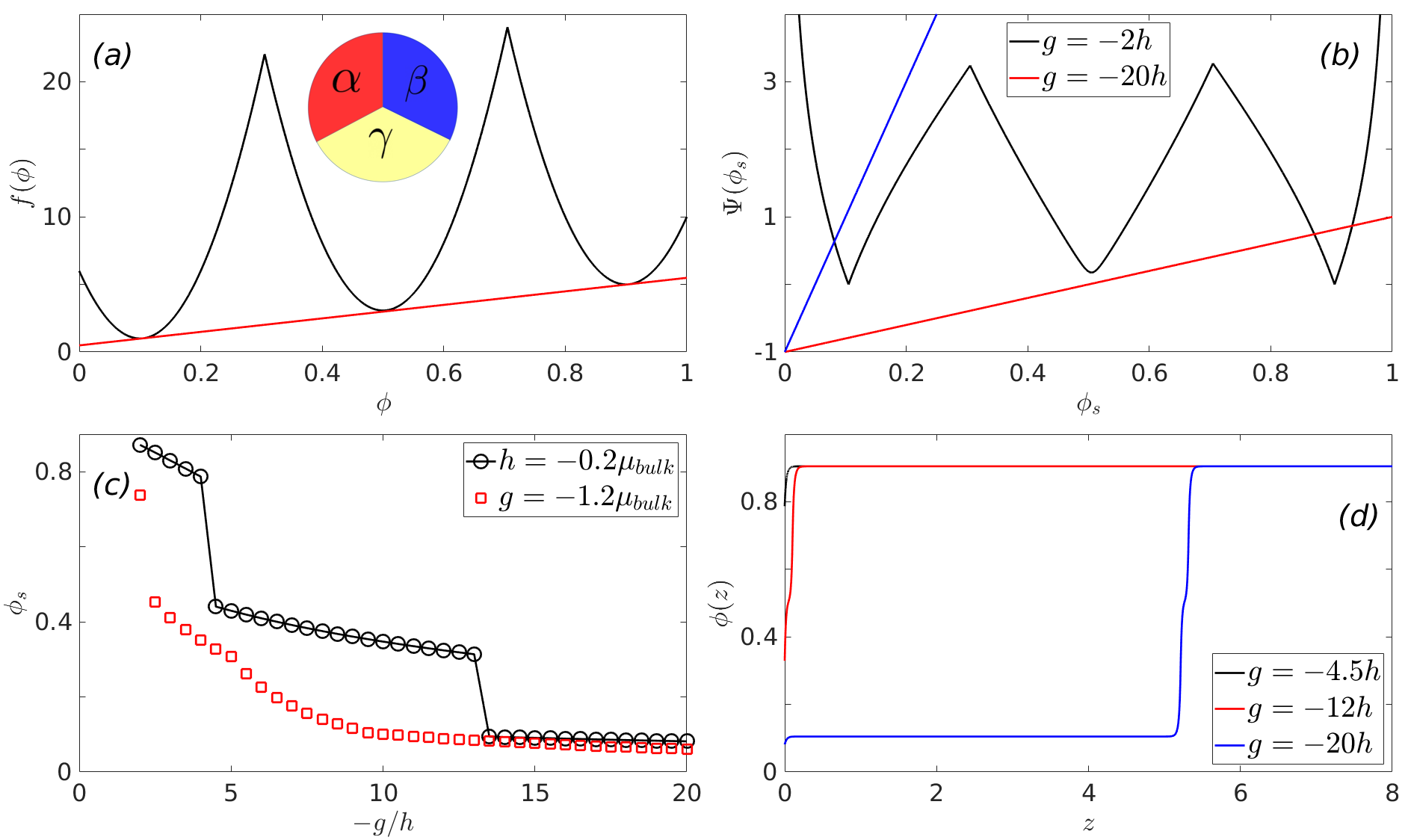} \\
\caption{The wetting thermodynamics as function of the $h$ and $g$ parameters, close to the triple point, where all the three phases coexist. Panel (a) shows the free energies (black line) and the common tangent in red, panel (b) shows the Cahn construction, when $h = -0.2  \mu_{bulk}$ and corresponding to the smallest and the largest g values considered. Panel (c) shows the surface fractions as a function of parameter g, for $h = -0.2 \mu_{bulk}$ (black line) and $h = -1.2 \mu_{bulk}$ (red line). Panel (d) shows the order parameter profiles for three values of g corresponding to $h = -0.2 \mu_{bulk}$.}
\label{h_g_phase_diag_triple_pt}
\end{center}
\end{figure}

In the final set of calculations with the model three-minimum free energy we compute the wetting phase diagram when the system is close to the triple point (where all three phases coexist) and vary the parameters, $h$ and $g$, which parametrises bare surface free-energy, $\Phi(\phi_{s})$. Panel (a) of Figure \ref{h_g_phase_diag_triple_pt} shows the triple-minimum free energy close to triple point and a common tangent showing the coexistence of all the three phases. In these calculations the value of the parameter $g$ is varied systematically form $g_{min} = -2h$ to $g_{max} = -20h$. The value of $h$ is again varied between $h_{min} = -0.2 \mu_{bulk}$ to $h_{max} = -1.2 \mu_{bulk}$, where $\mu_{bulk}$ is the slope of the common tangent in panel (a). The corresponding Cahn plots for the lines $\frac{d \Phi (\phi_s)}{d \phi_s}$, with the smallest and largest slopes are shown in panel (b), where $h = -0.2 \mu_{bulk}$. The surface lines correspond to  $\frac{d \Phi (\phi_s)}{d \phi_s} = h + g \phi_{s}$, thus h is the intercept of the surface line and $g$ is its slope. In panel (c) of Figure \ref{h_g_phase_diag_triple_pt} we observe that at low absolute value of the parameter $h$, we observe two first order transitions (black line), for the surface fraction as a function of the parameter $g$, of which the first transition occurring at a value of $-g/h \sim 5$ is between two partially wet states, whereas the transition occurring at $-g/h \sim 13$ is a transition between partial to complete wetting states. 

Upon increasing the absolute value of $h$ (red line), the first order transition at occurring at higher value of $g$, transforms to a continuous transition and also the jump in the surface fraction, $\phi_{s}$, occurring at low $g/h$, also decreases. The first order transitions occur when the line corresponding to the derivative of the surface free-energy, $\frac{d \Phi (\phi_s)}{d \phi_s}$, cuts the curve $2 \sqrt{\Delta f'(\phi_s) k(\phi_s)}$ simultaneously at three values of $\phi_{s}$ and this only happens when the slope of the $\frac{d \Phi (\phi_s)}{d \phi_s}$ line is small as in panel (b). When the magnitude of $h$ increases, the $\frac{d \Phi (\phi_s)}{d \phi_s}$ line never cuts the curve described by $2 \sqrt{\Delta f'(\phi_s) k(\phi_s)}$ simultaneously at three points and transitions tuned by varying parameter $g$ become continuous in nature \cite{p:degennes.rmp.v57.p827.y1985}. Panel (d) of Figure \ref{h_g_phase_diag_triple_pt} shows the order parameter profiles for the three values of g, when h is set to -0.2 $\mu_{bulk}$. At the highest absolute value $g$ (blue line) we observe a complete wetting of the surface by the $\alpha$ phase. As the system is close to the triple point and as the common tangent simultaneously passes through all the three minima, the $\alpha$ phase at the surface is wet by the $\beta$ phase and finally the $\gamma$ phase emerges deep in the bulk. for lower values of the parameter $g \sim -12 h$ (red line), one observes the $\beta$ phase at the surface, which then leads to the $\gamma$ phase in the bulk.

\section{Wetting of Polymer Dispersed Liquid Crystal mixtures:}
We apply the methodology developed for studying the wetting thermodynamics of a generic three-phase system to the case of polymer dispersed liquid crystals. Here, we use as an example a model of PDLC previously studied by Matsuyama \textit{et al.} \cite{p:matsuyama.epje.v9.p79.y2002,p.matsuyama.epje.v9.p89.y2002} for describing the bulk thermodynamics of a mixture of polymers and nematogens. A Flory-Huggins type free energy of the mixture, depending on two order parameters, is given by the free energy,

\begin{equation}
    f(\phi,S) = f_{iso}(\phi) + \nu \phi^{2} f_{nem}(S),
\end{equation}

where $\phi$ is the composition of the nematic component, $(1 - \phi)$ is the composition of the polymer. The term $f_{iso}(\phi)$ is the Flory-Huggins like isotropic part of the free-energy, given by,

\begin{equation}
    f_{iso}(\phi) = \frac{1-\phi}{n_{P}} \ln (1 - \phi) + \frac{\phi}{n_{l}} \ln \phi + \chi \phi (1 - \phi),
\end{equation}
where $n_{P}$ is the length of the polymer, $n_{l}$ is the length of the nematogens and $\chi$ is the Flory-Huggins parameter controlling the thermodynamics of mixing. The nematic part of the free energy is given by $f_{nem}(S)$, with $S$ denoting the nematic order parameter, is given by,

\begin{equation}
    f_{nem}(S) = \frac{1}{2} \left [(1 - \frac{\eta}{3}) S^{2} - \frac{\eta}{9} S^{3} + \frac{\eta}{6} S^{4} \right].
    \label{eq:nematic_free_energy}
\end{equation}

In Eq.~\ref{eq:nematic_free_energy} $\eta$ is a factor dependent of the local nematic density $\phi$, which couples the polymeric and the liquid crystalline degrees of freedom and is given by $\eta$ = $n_{l} \nu \phi$. The parameter $\nu$ controls the isotropic to nematic transition and is given by 
\begin{equation}
    \nu = \frac{2.7}{n_{l}} \left( \frac{T_{NI}}{T} \right).
\end{equation}
As a result of this, $\eta$ is given by,
\begin{equation}
\eta = 2.7 \left( \frac{T_{NI}}{T} \right) \phi
\end{equation}
Similarly, $\chi$, the parameter controlling the phase separation is given by,

\begin{equation}
    \chi = \frac{1}{n_{P}} \left( \frac{T_{c}}{T} \right)
 \end{equation}
 
Thermodynamic equilibrium necessitates the minimisation of the total free energy, which we achieve in two steps : first we minimise the nematic part of the free energy and obtain a value of the nematic order parameter $S$ (which is a function of $\eta$, which inturn is a function of $\phi$). This order parameter $S$ is then substituted back into the free energy, which now becomes a renormalised function of $\phi$.
 
Upon minimising $f_{nem}(S)$, we get the following equation for the non-zero roots,
 
 \begin{equation}
     \frac{2 \eta}{3} S^{2} - \frac{\eta}{3}S + (1 - \frac{\eta}{3}) = 0
 \end{equation}
 This equation has two roots, of which the positive (below $T_{NI}$ only the positive root contributes) is given by
 \begin{equation}
     S_{+} = \frac{\eta/3 + \sqrt{\eta^{2} - 8\eta/3}}{(4 \eta/3)}
 \end{equation}
 The order parameter $S_{+}$ is now substituted back into the expression of $f(\phi, S)$ resulting in a renormalized free energy which is a function of $\phi$ alone. The thermodynamics of this model is derived form this modified free energy.

\begin{figure}[h]
\begin{center}
\includegraphics[width=8.5cm]{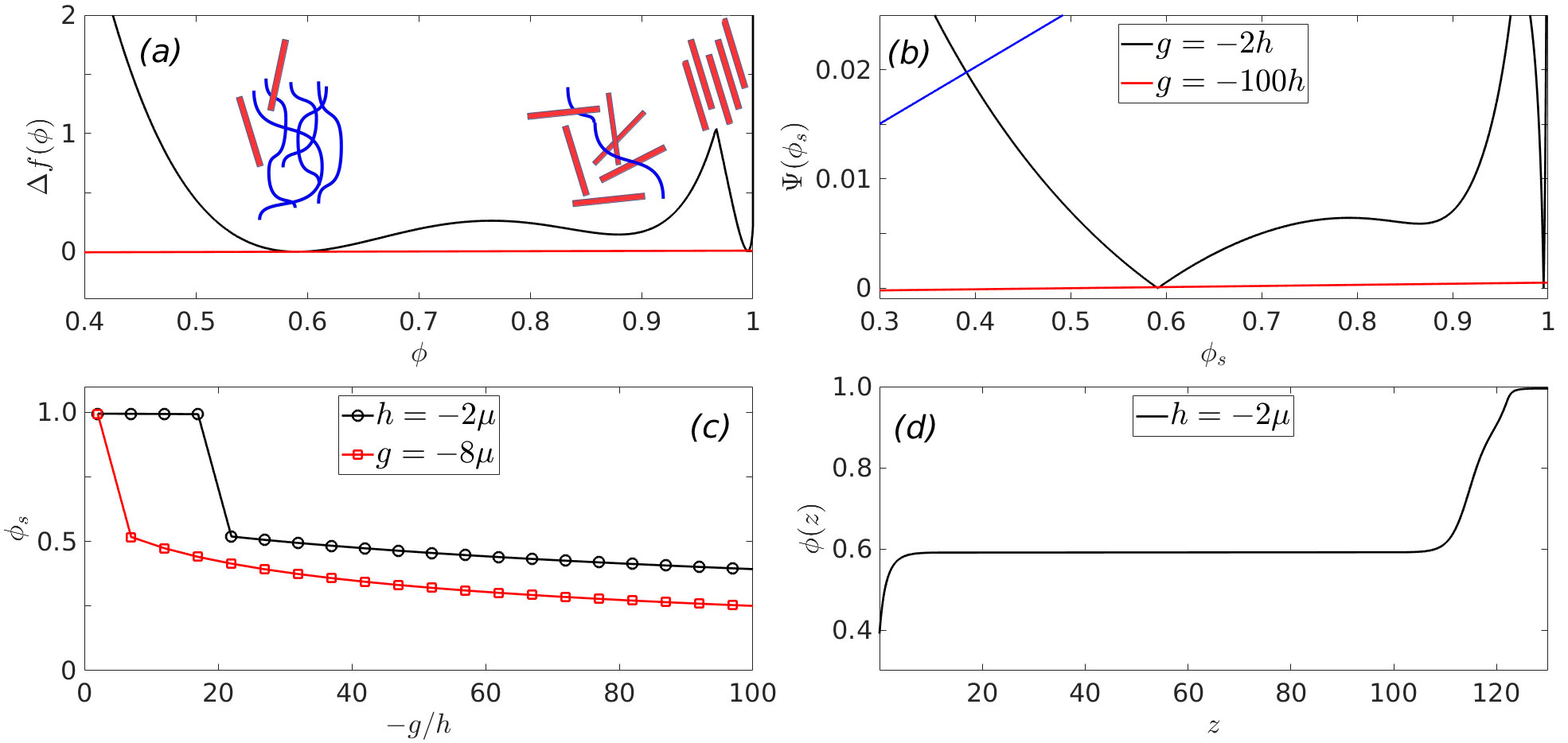}
\caption{The renormalised free-energy is shown in panel (a) (after the minimisation has been performed on the nematic part of the free-energy) as a function of the nematic volume fraction $\phi$, showing the low density isotropic phases $I_{1}$, the high density isotropic phase $I_{2}$ and the nematic phase $N$. Panel (b) shows the Cahn construction, with the surface lines shown for the minimum and maximum $g$ corresponding to $h = -2 \mu$. Panel (c) shows the variation of the surface fraction as a function of the parameter $g$ ($h = -2\mu$ is shown in black, while $h = -20 \mu$ is shown in red). Panel (d) shows the profile of the order parameter corresponding to the surface line shown in blue in panel (b).}
\label{PDLC_wetting}
\end{center}
\end{figure}

We study a system for which $n_{p}$ = 20, $n_{l}$ = 2 and $\nu/\chi$ = 3.1 and we are close to the triple point of the system at $\tau$ = 0.969, where the two isotropic phases, $I_{1}$ and $I_{2}$ and the nematic phase $N$ are in coexistence. The bulk free energy or the free-energy difference of the system with respect to an initially homogeneous state, which enters the wetting calculation is given by,
\begin{equation}
    \Delta f(\phi,S) = f(\phi,S) - f(\phi_{0},0) - (\phi - \phi_{0})(\partial f/\partial \phi)_{\phi = \phi_{0}} 
\end{equation}
where $\phi_{0}$ refers to the order-parameter of the initially homogeneous system and its value is taken as 0.6 in the subsequent calculations. It is also assumed that the surface prefers the 
polymeric component characterised by low value of the order parameter $\phi$. This free energy is shown in panel (a) of Figure \ref{PDLC_wetting}, which has three minima around $\phi$ $\sim$ 0.6 (isotropic), 0.88 (isotropic) and 0.99 (nematic). The parameters describing the surface interaction energy, $\Phi (\phi_s) = h\phi_{s} + \frac{1}{2}g\phi_{s}^2$, are the following : 
$g$ is varied between $-2h$ and $-100h$, where $h$ is varied between -2$\mu$ and -8$\mu$, where $\mu$ is the slope of the common tangent between the minima at $\phi = 0.6$ and the one at $\phi = 0.88$, in panel (a) of Figure \ref{PDLC_wetting}. We observe qualitatively similar features in wetting behaviour to our previously discussed model three-minimum free energy. Panel (b) shows the Cahn construction for the surface lines shown for the minimum and maximum $g$ corresponding to $h = -2 \mu$. We observe in panel (c) that at low absolute value of the parameter $h$, the surface fraction undergoes a first order transitions (black line), as a function of the parameter $g$, while at higher absolute values of parameter h one observes continuous transition in the surface fraction (red line). Panel (d) shows the profile of the order parameter corresponding to the surface line shown in blue in panel (b).

\textit{Conclusions:}
We discuss a mean-field theory for the thermodynamics of wetting in complex mixtures, where there are three minimum in the bulk free-energy landscape when exposed to a surface, whic prefers one of the components. Such a free-energy landscape can arise in a variety of complex mixtures like polymer nematic mixtures, ternary amphiphiles, polymer-colloid mixtures or metallic alloys. Interaction with the external surface are accounted via local potentials. We apply the Cahn-Landau-De Gennes mean field theory to understand the wetting thermodynamics of such a system as we sytematically vary the height of the central minimum and we find that the surface tension decreases monotonically with the height of this minimum, when it is unstable. As the central minimum becomes stable the phase diagram bifurcates and we observe a non-monotonic dependence of the surface tension on the stability of the central minimum, in one of the branches, which is associated with a complete to partial wetting transition. In the other branch we observe a monotonic increase in surface tension with an increasing stability of the central minimum. Close to the triple point, the wetting phase diagram computed by varying the bare surface energy parameters, $h$ and $g$, yields two first order transitions in the surface fraction as a function $g$, for low values of the parameter $h$. Upon increasing the absolute values of $h$, we observe that the first order transition in surface fractions give way to continuous transitions. A geometric understanding of these phenomena is discussed. Finally we present the wetting calculations for a polymer-nematic mixture, whose free energy actually has a three-minimum structure and show that the qualitative results obtained for our generic three-minimum free energy also holds for the polymer-nematic mixture.

We hope that our theoretical work will prompt experimental studies in understanding wetting and phase behavior of biopolymer solutions in cellular environments. Our results are also applicable to polymer dispersed liquid crystals (PDLC) which are an important class of materials with applications ranging from novel bulk phenomena in electro-optic devices \cite{bronnikov2013polymer} to very rich and unique surface phenomena like tunable surface roughness \cite{liu2015reverse} and electric-field driven meso-patterning on soft surfaces \cite{roy2019electrodynamic,dhara2018transition}.

\textit{Acknowledgements:}
BC and BM acknowledge funding support from EPSRC via grant EP/P07864/1, and P $\&$ G, Akzo-Nobel, and Mondelez Intl. Plc. We would also like to acknowledge S. Biswas for help in preparing the figures.


\end{document}